\documentclass[%
 preprint,
 amsmath,amssymb,
aps,
superscriptaddress
]{revtex4-1}
\usepackage{graphicx}% Include figure files
\usepackage{dcolumn}% Align table columns on decimal point
\usepackage{bm}% bold math
\usepackage{here}
\usepackage{braket}
\usepackage{url}

\usepackage{color}

\newcommand*{\NCO}[0]{NiCo$_2$O$_4$}

\begin{document}
\title{Spin reorientation in tetragonally distorted spinel oxide \NCO\  epitaxial films}

\author{Hiroki Koizumi} 
\affiliation{Department of Applied Physics, University of Tsukuba, Tsukuba, Ibaraki 305-8573, Japan}

\author{Ikumi Suzuki} 
\affiliation{Institute for Chemical Research, Kyoto University, Uji, Kyoto 611-0011, Japan}

\author{Daisuke Kan} 
\affiliation{Institute for Chemical Research, Kyoto University, Uji, Kyoto 611-0011, Japan}

\author{Jun-ichiro Inoue} 
\affiliation{Department of Applied Physics, University of Tsukuba, Tsukuba, Ibaraki 305-8573, Japan}

\author{Yusuke Wakabayashi} 
\affiliation{Department of Physics, Tohoku University,  Sendai 980-8578, Japan}

\author{Yuichi Shimakawa} 
\affiliation{Institute for Chemical Research, Kyoto University, Uji, Kyoto 611-0011, Japan}

\author{Hideto Yanagihara}
\affiliation{Department of Applied Physics, University of Tsukuba, Tsukuba, Ibaraki 305-8573, Japan}
\affiliation{Tsukuba Research Center for Energy Materials Science (TREMS), University of Tsukuba, Tsukuba, Ibaraki 305-8573, Japan}

\begin{abstract}
We experimentally investigated the magnetic properties of \NCO\ epitaxial films known to be conductive oxides with perpendicular magnetic anisotropy (PMA) at room temperature.
Both magneto-torque and magnetization measurements at various temperatures provide clear experimental evidence of the spin reorientation transition at which the MA changes from PMA to easy-cone magnetic anisotropy (ECMA) at a certain temperature ($T_{\rm{SR}}$).
ECMA was commonly observed in films grown by pulsed laser deposition and reactive radio frequency magnetron sputtering, although $T_{\mathrm{SR}}$ is dependent on the growth method as well as the conditions. 
The cone angles measured from  the $c$-axis increased successively at $T_{\mathrm{SR}}$ and approached a maximum of 45-50 degrees at the lowest measurement temperature of 5 K.
Calculation with the cluster model suggests that the Ni$^{3+}$ ions occupying the $T_d$ site could be the origin of the ECMA.
Both the magnetic properties and the results of the calculation based on the cluster model indicate that the ECMA is attributable to the cation anti-site distribution of Ni$^{3+}$, which is possibly formed during the growth process of the thin films. 
\end{abstract}

\maketitle

\section{Introduction}
\NCO (NCO) is a conductive inverse spinel oxide ($Fd\bar{3}m$) that exhibits  ferrimagnetism with a Curie temperature as high as $T_C \approx 400$ K \cite{Battle1979, MARCO2000}.
Owing to the mixed valence of its cations, NCO can be expressed in terms of the following chemical formula: $(\mathrm{Co}^{2+}_x\mathrm{Co}^{3+}_{1-x})_{\rm tet}[\mathrm{Co}^{3+}\mathrm{Ni}^{2+}_{1-x}\mathrm{Ni}^{3+}_x]_{\rm oct}\mathrm{O}_4^{2-}$ ($0<x<1$), where the subscripts ``tet'' and ``oct'' represent the tetrahedral ($T_d$) and octahedral ($O_h$) sites, respectively.
The $T_d$ sites are occupied by $\mathrm{Co}^{2+}$ ($d^7$; $S=3/2$) and $\mathrm{Co}^{3+}$ ($d^6$; $S=2$).
On the other hand, the $O_h$ sites are occupied by $\mathrm{Ni}^{2+}$ ($d^8$; $S=1$), $\mathrm{Ni}^{3+}$ ($d^7$; $S=1/2$), and $\mathrm{Co}^{3+}$ ($d^6$; $S=0$).
From the viewpoint of the number of $d$-electrons at each site, the nominal saturation magnetization is 2 $\mu_B$/f.u. irrespective of the value of $x$, as long as the $T_d$ sites are only occupied by Co ions.

The mixed valence properties of NCO confer unique characteristics upon this transition metal oxide, such as relatively high electrical conductivity \cite{Silwal2013, Ndione2014, Bitla2015, Zhen2018}, and electrochemical activity \cite{Wang2019}. 
In particular, a large tunnel magnetoresistance effect has recently been reported in magnetic tunnel junctions fabricated using NCO electrodes, supporting the half-metallic nature\cite{Ndione2014,Wang2019,Shen2020_TMR}.
In addition, NCO(001) films grown on MgAl$_2$O$_4$ (MAO) substrates exhibited perpendicular magnetic anisotropy (PMA) \cite{Chen2019, Kan2020, Xue2020, Koizumi2021} of $K^{\rm{eff}}_{u} \approx  0.3$ MJ/m$^3$ at room temperature. The fact that both the half-metallicity and PMA are realized at once indicates that the NCO films possess good potential for application in spintronic devices. 
Mellinger {\it et al.} remarked that the PMA is due to the magneto-elastic effect, which is provided a semi-quantitative explanation for its attribution of the PMA to the compressive strain of $\sim -0.3\%$ to which Co at the $T_d$ sites is subjected as a result of epitaxial distortion \cite{Mellinger2020}.

In this paper, we report that NCO films have spin reorientation (SR) transitions at low temperatures, that is, the magnetic anisotropy (MA) of the films changes from PMA to easy-cone magnetic anisotropy (ECMA). This suggests that the higher order term of the uniaxial MA must be considered at least in the low-temperature region.
To elucidate the origin of the ECMA, we also performed tight-binding calculations for Co-O and Ni-O clusters of either the divalent or trivalent states at both the $T_d$ and $O_h$ sites. The results showed that the observed ECMA predominantly originated from the Ni$^{3+}$ that occupy the $T_d$ sites, which is the anti-site disorder of cation distribution.

This paper is organized as follows. 
Experimental details are provided in Sec. \ref{sec2}. 
Section \ref{sec3} is devoted to the experimental results, numerical calculation, and discussion, and the final section provides a summary of the work.

\section{Experiments}\label{sec2}
We prepared three epitaxial NCO films grown on single-crystal MAO(001) substrates by reactive radio frequency magnetron sputtering (ES-250MB, Eiko Engineering Co., Ltd.)\cite{Koizumi2019prm,koizumi2019prb} or pulsed laser deposition (Pascal Co., Ltd.).
In the sputtering process, we used a 2-inch alloy target with a nominal composition of $\mathrm{Ni:Co} = 1:2$. 
The growth conditions of the NCO thin films by sputtering were as follows: the Ar and O$_2$ flow rates were 10 and 2.5 sccm, respectively, and the input power was 100 W.
The process temperature and working pressure were 300$^\circ$C and 1.3 Pa, respectively. 
The film thickness was determined to be 50 nm by X-ray  reflectivity measurement.
Hereafter, we refer to this NCO film as Sample $S$.
On the other hand, in the case of the PLD process, we used a NiCo$_2$O$_x$ ceramic target, which was ablated by a 6 Hz KrF excimer laser ($\lambda  = 248$ nm) with a laser spot density of 1.2 J/cm$^2$, at a fixed substrate temperature of 315$^\circ$C, and under various oxygen partial pressures.
Detailed process conditions can be found in Refs. \onlinecite{Shen2020, Shen2020_TMR, Suzuki2020}.
The NCO films prepared at an oxygen pressure of $P_\mathrm{O_2} = 10$ and 100 mTorr  are denoted as Samples $P_{10}$ and $P_{100}$, respectively. 
The film thicknesses of both Samples $P_{10}$ and $P_{100}$ were 30 nm.

Thin film structures were characterized by X-ray diffraction (XRD).
Magnetic hysteresis (MH) loops and magnetic anisotropy constants were measured using a vibrating sample magnetometer (VSM) and magneto-torque meter, respectively. 
Both magnetic measurements were performed using a physical property measurement system (PPMS, Quantum Design).

\section{Results and discussion}\label{sec3}

\subsection{Structural characterization}
\begin{figure}[htb]
\begin{center} 
\includegraphics[keepaspectratio , width=8.6cm]{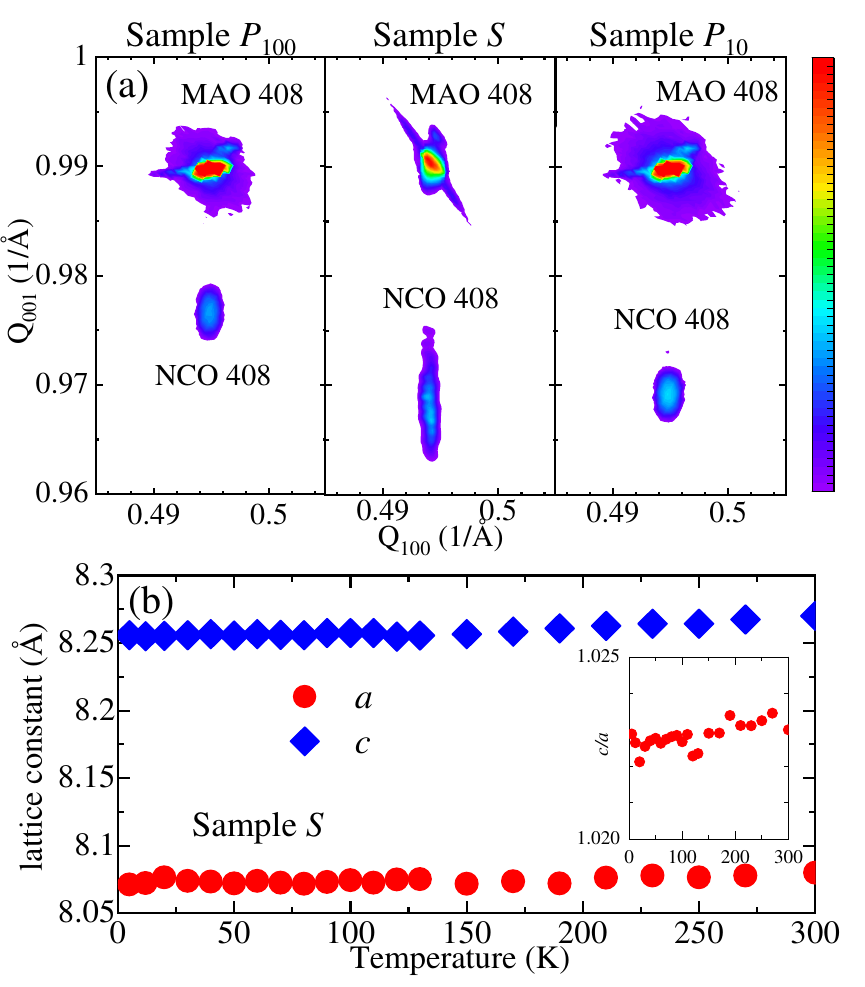}
\caption{(a) RSMs around the NCO$(4, 0, 8)$ and MAO$(4, 0, 8)$ reflections. (b) Temperature dependence of the lattice constant evaluated from the NCO$(4, 0, 12)$ Bragg reflection in Sample $S$. The inset shows the temperature dependence of the $c/a$ ratio.}
\label{XRD}
\end{center} 
\end{figure}

Figure \ref{XRD} (a) shows the reciprocal space map (RSM) around the MAO$(4, 0, 8)$ Bragg reflections for each sample. 
All three films are tetragonally distorted by the MAO substrate. 
The in-plane lattice constant seems to be constrained by the MAO substrate as  $a = 8.08$ \AA, whereas the out-of-plane ($c$) constant varies from sample to sample, namely  $c = 8.19$ \AA\ ($c/a=1.013$), 8.27 \AA\ ($c/a=1.023$),  and  8.25 \AA\ ($c/a=1.021$) for Samples $P_{100}$, $S$, and $P_{10}$, respectively. 
The lattice parameter along the $c$-axis varies from sample to sample, suggesting that the cation distribution varies according to the growth conditions of the films \cite{Shen2020}.

We also measured the temperature dependence of the lattice parameters for Sample $S$, as shown in Fig. \ref{XRD} (b). 
Both the in-plane and out-of-plane lattice constants $a$ and $c$, determined from the NCO$(4, 0, 12)$ Bragg reflections, are less temperature dependent and are therefore similar to that of the $c/a$ ratio.

\subsection{Magnetic properties}
\begin{figure}[htb]
\begin{center}
\includegraphics[keepaspectratio , width=8.6cm]{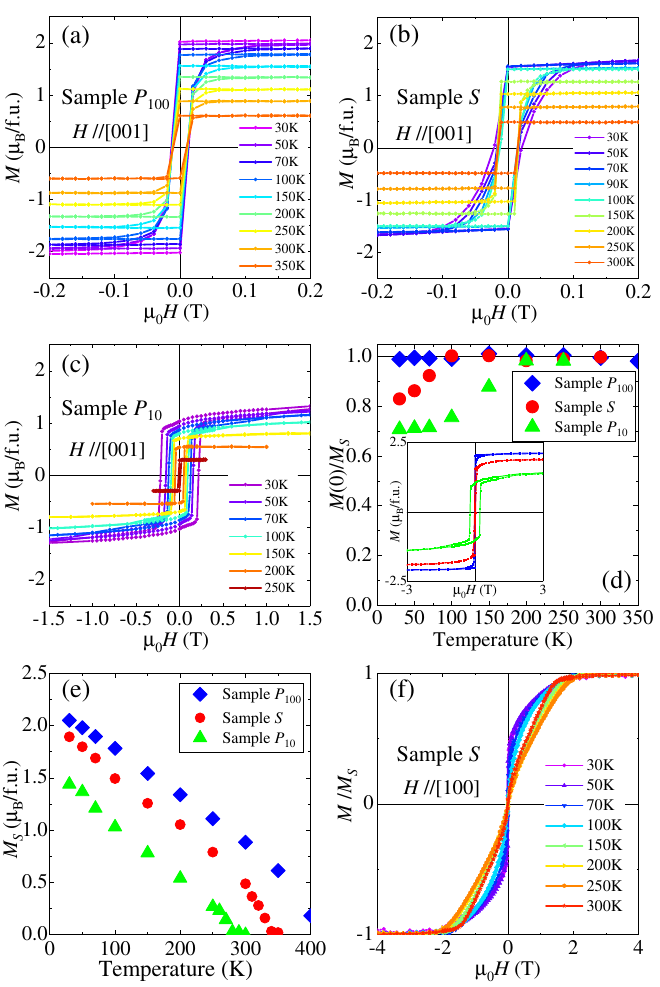}
\caption{ $MH$ loops of NCO film with $\mu _0 H \perp$film plane of (a) Sample $P_{100}$, (b) Sample $S$, and (c) Sample  $P_{10}$. (d) Temperature dependence of the squareness ratio in each sample (inset: MH loops for the three samples at 30 K). (e) Temperature dependence of $M_S$ determined from VSM results. (f) $\mu _0 H \parallel$ film plane of Sample $S$.}
\label{VSM}
\end{center} 
\end{figure}

Figure \ref{VSM} (a)-(c) shows the temperature dependence of $MH$ loops in the presence of a magnetic field applied perpendicular to the film plane.
All films exhibited square-shaped hysteresis loops at temperatures above 250 K, whereas the coercivity varies markedly among the samples.
Moreover, with decreasing temperature, a stronger field is needed to saturate the magnetization at low temperatures in Samples $P_{10}$ and $S$, meaning that the squareness of the hysteresis is reduced.
The squareness ratio starts to decrease at a certain temperature, as shown in Fig. \ref{VSM} (d).
For example, the squareness ratio of Sample $P_{10}$ decreases from 200 K and almost constant below 70 K, and that of Sample $S$ starts decreasing at approximately 70 K.
For Sample $P_{100}$, however, the squareness ratio appears to be close to 1 at all measured temperatures.
The squareness ratios at 30 K, the lowest temperature at which we performed  measurements, are 0.99, 0.81, and 0.73 for $P_{100}$, $S$ and $P_{10}$, respectively (see the inset of Fig. \ref{VSM} (d)).
The squareness at lower temperatures is obviously smaller for Samples $S$ and $P_{10}$, although the $M(T)$ curves of none of the three films exhibited any anomalies.

Figure \ref{VSM} (e) shows the temperature dependence of the saturation magnetization $M_S$, with higher values being observed for samples with higher Curie temperature ($T_C$).
Because $T_C$ and $M_S$ are simply associated with the strength of the exchange interaction and the difference in the magnetic moments between the $O_h$ and $T_d$ sites, respectively, the clear positive correlation between $T_C$ and $M_S$ indicates that the distribution of both Co and Ni at both the $T_d$ and $O_h$ sites changes with the growth conditions.
In fact, according to a recent structural analysis of NCO films, the amount of cation distribution defects or anti-site defects characteristic of NCO films is non-negligible \cite{Shen2020}.

Figure \ref{VSM} (f) shows the $MH$ loops of Sample $S$ with the magnetic field parallel to the film plane.
These results reveal that the $MH$ loops became rounded and the area corresponding to $\int_0^{M_s} \mu_0H(M) dM$ tended to be smaller as the temperature decreases.
Considering the lowest order term of MA, these results indicate that the effective MA decreases at low temperatures.
Note that this does not occur in conventional ferromagnetic or ferrimagnetic materials; therefore, a detailed evaluation of MA must be carried out, especially at low temperatures.

\begin{figure}[htb]
\begin{center} 
\includegraphics[keepaspectratio , width=8.6cm]{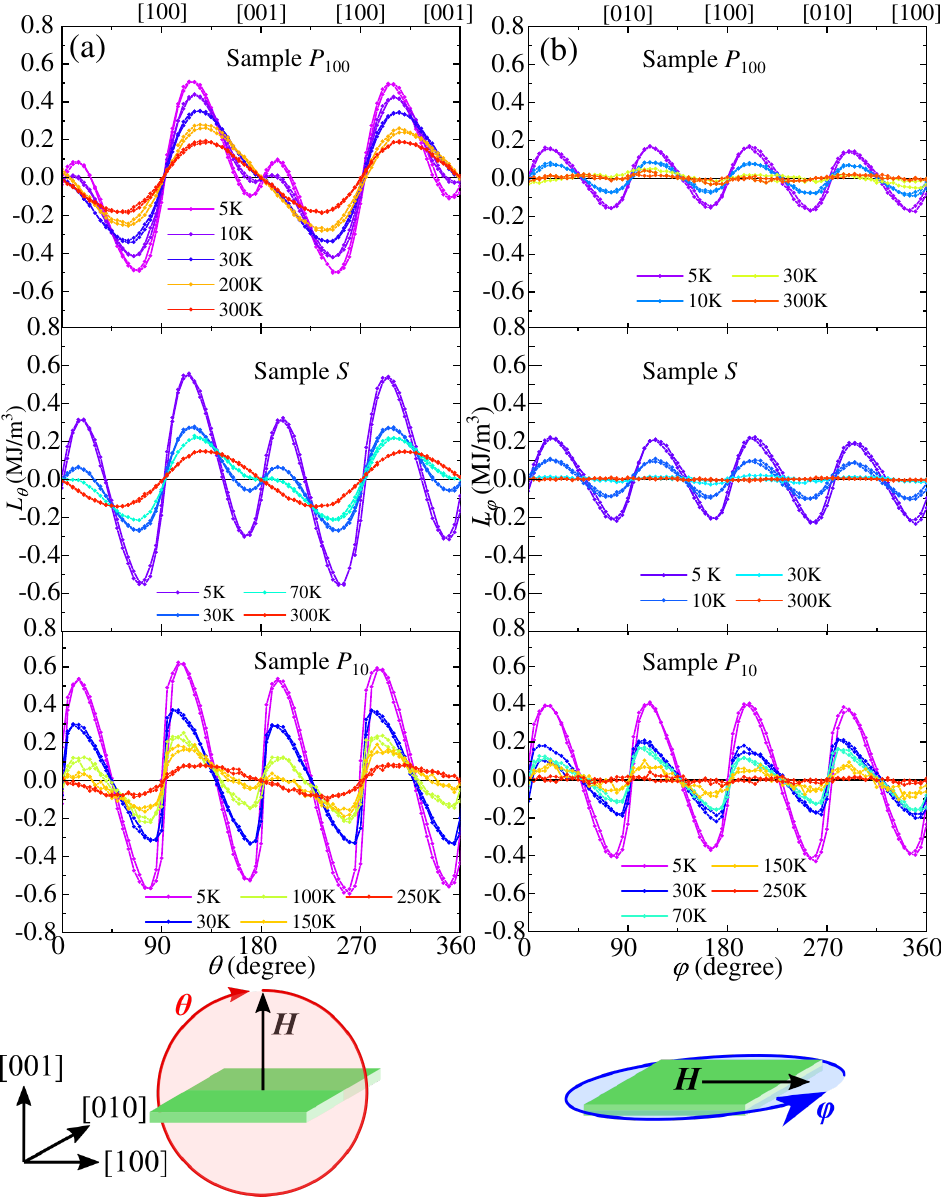}
\caption{ Magneto-torque curves of (a)$L_\theta$ ($ac$-plane), and (b)$L_\varphi$ ($ab$-plane) for the three samples at various temperatures. }
\label{Torque}
\end{center} 
\end{figure}

The magneto-torque curves of the $ac$-plane (around the $b$-axis, $L_\theta $) and of the $ab$-plane (around the $c$-axis; $L_\phi$) from room temperature to 5 K  taken under $\mu _0 H = 9$ T are shown in Fig. \ref{Torque}.
As shown in Fig. \ref{Torque} (a), the shape of $L_\theta$ at high temperatures exhibit two-fold component  crossing zero with a negative slope at $\theta = 0$ and $180$ degree, indicating the preferential magnetization directions of the films correspond to the direction normal to the film plane, which is consistent with previous reports \cite{Kan2020,Mellinger2020}.
However, four-fold components appear in the torque curves with decreasing temperature.
The existence of four-fold components means that the higher-order terms in uniaxial MA are crucial for understanding the preferential magnetization directions \cite{Chikazumi}.

In the $L_\varphi$ curves shown in Fig. \ref{Torque} (b), on the other hand, the torque signals are observed only at low temperature and are weaker than those of $L_\theta$, meaning that the out-of-plane uniaxial components are dominant in the NCO films.

In tetragonal crystal systems, the magnetic anisotropy energy ($E_A$) and magneto-torque for the $ac$-plane ($L_\theta $) and $ab$-plane ($L_\varphi  $) can be written as follows \cite{Chikazumi,Coey}:
\begin{eqnarray}
E_A&=&
K_{u1} ^{\mathrm{eff}}\sin ^2 \theta 
+K_{u2}\sin ^4 \theta 
+K_3\sin ^4 \theta \sin ^2 2\varphi,
\label{eq1}
\\
L_\theta
&\equiv&
-\partial_\theta E_A 
=
-\left(
K_{u1} ^{\mathrm{eff}}+K_{u2}
\right)\sin 2\theta
+\frac{1}{2}K_{u2}\sin 4\theta, \label{eq2}
\\
L_\varphi
&\equiv&
-\partial_\varphi E_A
=
-K_3\sin 4\varphi , \label{eq3}
\end{eqnarray}
where $K_{u1} ^{\mathrm{eff}}$ and $K_{u2}$ are the uniaxial magnetic anisotropy constants for the second- and fourth-order terms, respectively, and $K_3$ is the cubic anisotropy constant. Further, $\theta $ and $\varphi$ represent the angles between the direction of the applied field and the $c$- and $a$-axes, respectively.
The observed $K_{u1} ^{\mathrm{eff}}$ comprises both magnetic anisotropy contributions from magnetocrystalline anisotropy $K_{u1}$ and shape anisotropy $\mu _0 M_S ^2/2 $, i.e., $K_{u1} ^{\mathrm{eff}}=K_{u1}-\mu _0 M_S ^2/2$.
The MA constants enable us to determine the shape of the preferential directions of the net magnetization \cite{Chikazumi}, and these results are provided in Table \ref{Shape}.
\begin{table}[htb]
\caption{Relation between the respective magnetic anisotropy constants and schematics of preferential magnetisation orientation.}
\label{Shape}
\begin{ruledtabular}
\begin{tabular}{cccccc}
 \multicolumn{3}{c|}{$\ \ \ \ \ \ \ \ \ \ \ \ \ \ \ \ K_{u1} ^{\mathrm{eff}}>0\ \ \ \ \ \ \ \ \ \ \ \ $} 
 &\multicolumn{3}{c}{$K_{u1} ^{\mathrm{eff}}<0$} \\ \hline 
\multicolumn{2}{c}{$K_{u1} ^{\mathrm{eff}}+K_{u2} >0$} & \multicolumn{2}{|c|}{$K_{u1} ^{\mathrm{eff}}+K_{u2} <0$} & \multicolumn{2}{c}{$K_{u1} ^{\mathrm{eff}}+2K_{u2} >0$}  \\ 
 \multicolumn{2}{c}{}& \multicolumn{2}{|c|}{$K_{u1} ^{\mathrm{eff}}+2K_{u2} <0$} &\multicolumn{2}{c}{}  \\\hline 
 \multicolumn{2}{c}{
\begin{minipage}{28mm}
  \centering
  \includegraphics[keepaspectratio , width=28mm]{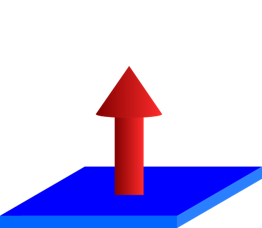}
\end{minipage} }&\multicolumn{2}{|c|}{
\begin{minipage}{28mm}
  \centering
  \includegraphics[keepaspectratio , width=28mm]{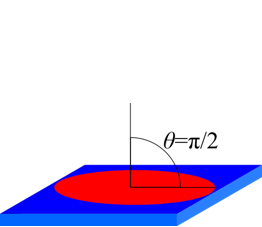}
\end{minipage} }&\multicolumn{2}{c}{
\begin{minipage}{28mm}
  \centering
  \includegraphics[keepaspectratio , width=28mm]{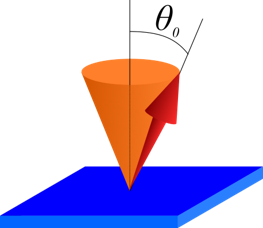}
\end{minipage}}\\ 
\multicolumn{2}{c}{Perpendicular} & \multicolumn{2}{|c|}{In-plane} & \multicolumn{2}{c}{Easy-cone} \\
\multicolumn{2}{c}{magnetic} & \multicolumn{2}{|c|}{magnetic} & \multicolumn{2}{c}{magnetic} \\
\multicolumn{2}{c}{anisotropy} & \multicolumn{2}{|c|}{anisotropy} & \multicolumn{2}{c}{anisotropy} \\
\multicolumn{2}{c}{(PMA)} & \multicolumn{2}{|c|}{}& \multicolumn{2}{c}{(ECMA)}  \\ 
\multicolumn{2}{c}{$T_{\mathrm{SR}}<T$} &  \multicolumn{2}{|c|}{}& \multicolumn{2}{c}{$T<T_{\mathrm{SR}}$} \\ 
\end{tabular}
\end{ruledtabular}
\end{table}

\begin{figure}[htb]
\begin{center} 
\includegraphics[keepaspectratio , width=8.6cm]{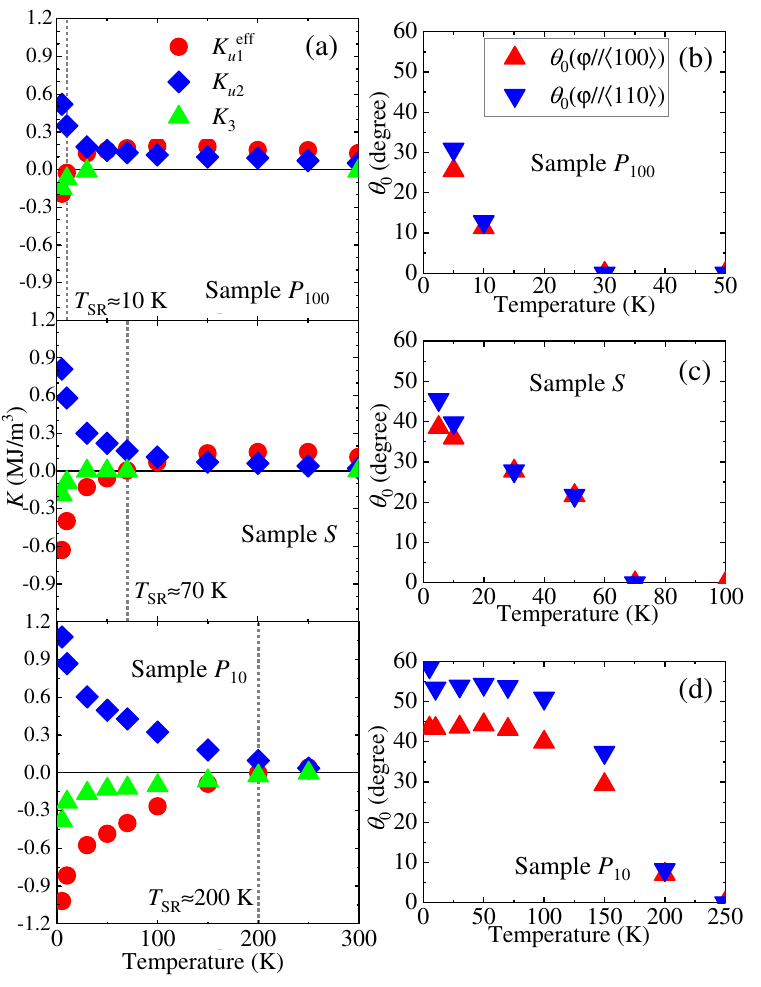}
\caption{ Temperature dependence (a) of the magnetic anisotropy energy in each sample and (b)–(d) of the cone angles for the two different $\varphi$ directions calculated from Eq. \ref{eq4}.
The vertical broken lines in (a) roughly indicate the temperatures where $K_{u1} ^{\mathrm{eff}}$  change their signs.}
\label{Anisotropy}
\end{center} 
\end{figure}

Figure \ref{Anisotropy} (a) shows the MA constants determined by fitting the torque curve using Eq. (\ref{eq2}), (\ref{eq3}).
The values of $K_{u1} ^{\mathrm{eff}}$ were positive at high temperatures and the sign changes to negative as the temperature decreases. 
In contrast, the values of $K_{u2}$ were positive at all temperatures.
Moreover, the absolute values of both $K_{u1} ^{\mathrm{eff}}$ and $K_{u2}$ increased suddenly and drastically when the temperature reached a certain point and this depended on samples.
On the other hand, $K_3$ is much weaker than the out-of-plane terms of both $K_{u1} ^{\mathrm{eff}}$ and $K_{u2}$.
Because $K_{u2}>0$ for all temperature ranges, if $K_{u1} ^{\mathrm{eff}}>0$ NCO exhibits PMA (see Table \ref{Shape}) \cite{Chikazumi}.
In contrast, when $K_{u1} ^{\mathrm{eff}}$ changes its sign from positive to negative, spin reorientation occurs, and the NCO exhibits ECMA. 
Hereafter, we refer to the temperature at which $K_{u1} ^{\mathrm{eff}}$ becomes $0$ as the spin-reorientation transition temperature ($T_{\mathrm{SR}}$).
The values of $T_{\mathrm{SR}}$ for all samples are listed in Table \ref{Summary}. 
It should also be noted that the temperature at which the squareness ratio begins to decrease is close to the value of $T_{\mathrm{SR}}$ determined by the torque measurements, indicating that the change in the squareness ratio is attributed to the change in MA. 
In addition, the cone angle determined from the magneto-torque is in good agreement with that estimated from the squareness ratio, as shown in Fig. \ref{VSM} (d).

By considering $K_3$, the cone angle $\theta _0 $ measured from the $c$-axis can be expressed as
\begin{eqnarray}
\theta _0 (\varphi)
&=&
\sqrt{
\frac{-K_{u1}^{\mathrm{eff}}}{2(K_{u2}+K_3 \sin ^2 2 \varphi)}. \label{eq4}
}
\end{eqnarray}
Figure \ref{Anisotropy} (b)-(d) show the temperature dependence of $\theta_0$, determined from Eq. (\ref{eq4}), for the three samples.
As the temperature decreases, $\theta_0$ continuously increases from $\theta_0 = 0$ to a finite value. 
We note that $\theta_0(T)$ for the MA of Co$_2$Y-hexaferrite was reported to exhibit similar behavior \cite{Smit1959}.

\begin{figure}[htb]
\begin{center} 
\includegraphics[keepaspectratio , width=5cm]{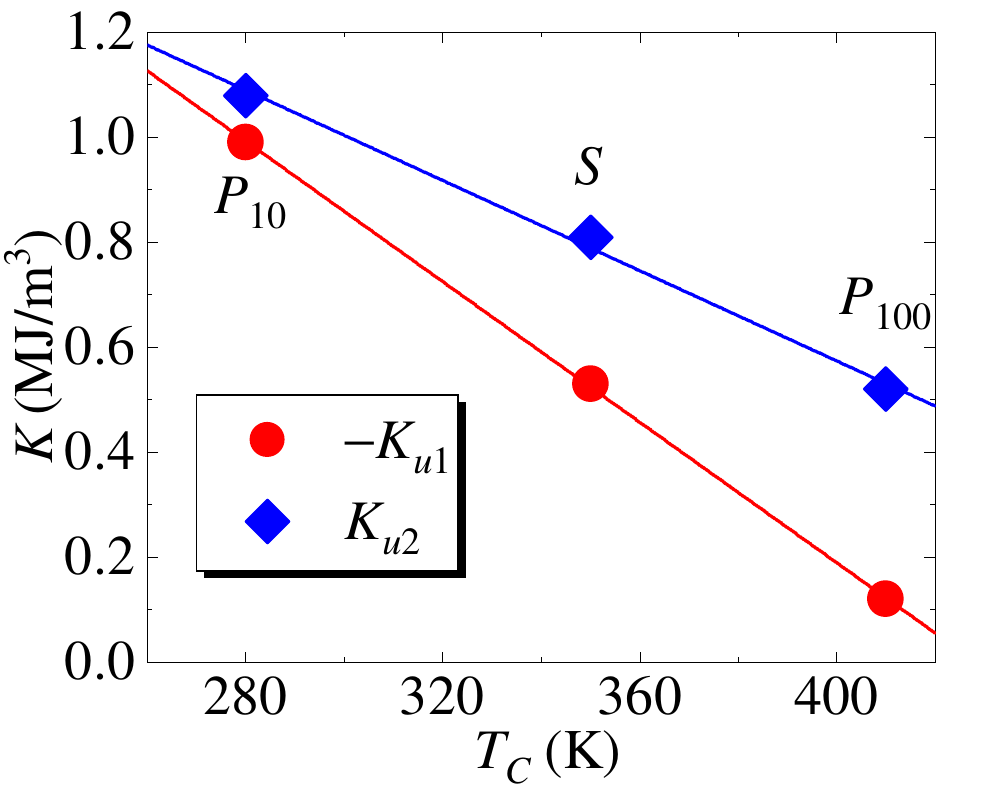}
\caption{ Curie temperature for the two respective magnetic anisotropy constants at 5 K for the three samples.}
\label{Sum}
\end{center} 
\end{figure}

The saturation magnetization extrapolated to 0 K ($M_0$), $T_C$, $T_{\mathrm{SR}}$, and the values of $K_u$ for each sample are summarized in Table \ref{Summary}.
The films with higher $T_C$ exhibit larger $M_0$ values and lower $T_{\mathrm{SR}}$.
Moreover, the MA constants seem to be linearly correlated with $T_C$, as shown in Fig. \ref{Sum}.
On the other hand, at approximately room temperature, the sample with higher $T_C$ exhibits stronger PMA, as shown in Fig. \ref{Anisotropy}.
According to a previous study \cite{Shen2020}, $T_C$ is higher for NCO films with less anti-site disorder.
In contrast, our experimental results suggest that samples with lower $T_C$ have stronger ECMA and higher $T_{\mathrm{SR}}$.
Therefore, ECMA seems to be associated with samples with off-stoichiometric compositions or anti-site distributions \cite{Shen2020}, which is consistent with the fact that no sign of a non-collinear spin structure was apparent even at 4.2 K in a stoichiometric powder sample of NCO by neutron diffraction \cite{Battle1979}.
Although $T_{\rm{SR}}$ varies from sample to sample, the ECMA emerges at low temperatures regardless of the sample preparation technique, indicating that ECMA generally emerges at low temperatures in tetragonally distorted NCO films.
\begin{table}[htb]
\caption{ Summary of the magnetic properties: saturation magnetization extrapolated to 0 K ($M_0$), magnetic anisotropy constants, $T_C$, and $T_{\mathrm{SR}}$.}
\label{Summary}
\begin{ruledtabular}
\begin{tabular}{cccc}
Sample 	&  $P_{100}$    & $S$	& $P_{10}$  \\ \hline
$M_0$ ($\mu _B$/f.u.)   & 2.2   & 2.0   & 1.6 \\
$K_{u1} $ (MJ/m$^3$)     & $-0.12$   & $-0.53$ & $-0.99$ \\
$K_{u2}$ (MJ/m$^3$)   & $0.52$     & $0.81$ & $1.08$\\
$T_C$ (K)               & $\sim 410$    &   $\sim 350$  & $\sim 280$ \\
$T_{\mathrm{SR}}$ (K)   &  $\sim 10$	& $\sim 70$     & $\sim 200$ \\	
\end{tabular}
\end{ruledtabular}
\end{table}

\subsection{Electron theory}
To understand the origin of the relatively large $K_{u2}$ component bearing ECMA, we performed a numerical calculation to evaluate the MA of each ion present in tetragonally distorted NCO within the framework of the cluster model \cite{Inoue_2014,Inoue2014AIP}.
Because of the presence of site-distribution disorders, we performed the calculations for eight different clusters using four ions, Co$^{2+}$, Co$^{3+}$, Ni$^{2+}$, and Ni$^{3+}$, which occupy either the $T_d$ or $O_h$ sites. 
Then, we estimated the total MA by considering plausible site distributions.

The electronic structure of the cluster was calculated using a tight-binding model for the 3$d$- and 2$p$-orbitals of the transition metal and oxygen ions, respectively, and by including the spin-orbit interaction (SOI).
We assumed that the 3$d$-states of the Co and Ni ions are fully spin-polarized, and that the ions have local moments.
The parameters of the inter-site $p$-$d$ hopping between the 3$d$-orbitals of a transition metal ion and the 2$p$-orbitals of the oxygen ions were obtained from Harrison's textbook \cite{Harrison1980,Inoue2015}.
The ground state energy was calculated by diagonalizing the Hamiltonian as a function of the magnetization direction. 
The lattice constant and Wyckoff position of NCO were obtained from a previous report \cite{Battle1979}.
The electron configurations and SOIs for each ion are summarized in Table \ref{configurations}.
\begin{table}[htb]
\caption{Summary of electron configurations of the ions. }
\label{configurations}
\begin{ruledtabular}
\begin{tabular}{cccccc}
Ion         & Co$^{2+}$     & Co$^{3+}$     & Ni$^{2+}$ & Ni$^{3+}$ & O$^{2-}$\\ \hline
Electron    & $3d^{7}$      & $3d^{6}$      & $3d^{8}$  & $3d^{7}$  & $2p^6$\\ 
SOI (meV)   & 20            & 30            & 30        & 30        & | \\ 
\end{tabular}
\end{ruledtabular}
\end{table}

In the calculation, we assumed that the Ni$^{3+}$ ions occupying the $O_h$ site are in a low-spin state of $S=1/2$ \cite{Bitla2015}. 
We excluded the calculation of Co$^{3+}$ in the $O_h$ site because of the low spin state of $S=0$.
To satisfy the above conditions, the Coulomb potential $U$ was set to be 0.05 eV for Ni in the $O_h$ site, whereas the potential for the others was $U=6.0$ eV.

\begin{figure}[htb]
\begin{center} 
\includegraphics[keepaspectratio , width=8.5cm]{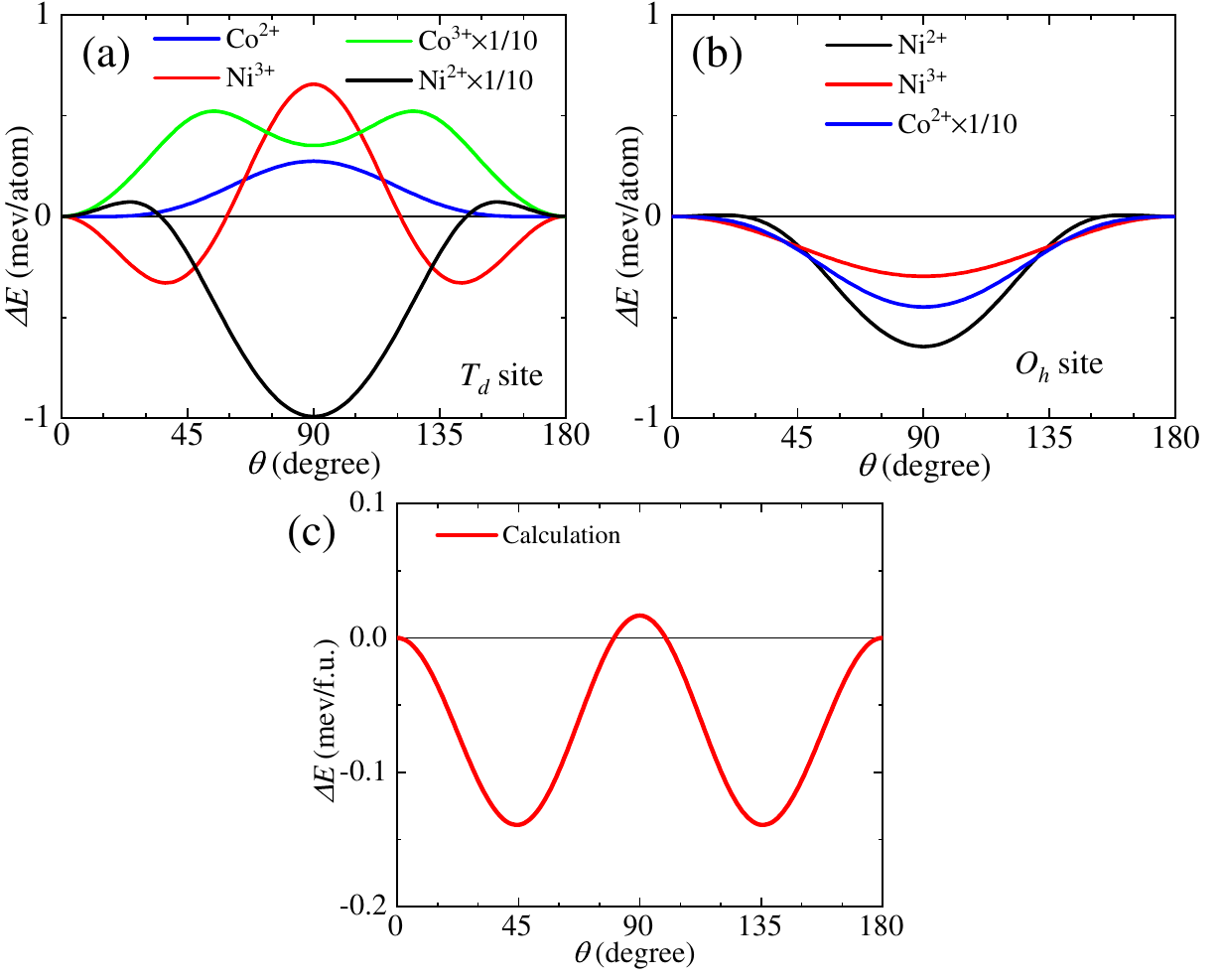}
\caption{Calculated dependence of the magnetic moment on the angle $\Delta E(\theta )$ at (a) $T_d$, and (b) $O_h$ sites.
(c) Calculated total anisotropy based on the adapted cation distribution in Table \ref{distribution}.
}
\label{Cluster}
\end{center} 
\end{figure}

Figure \ref{Cluster} (a) and (b) show the angle-dependent energy of the magnetic moment ($\Delta E$) at the $T_d$ and $O_h$ sites, respectively.
The preferential spin direction of Co$^{3+}$ at the $T_d$ site is parallel to the $c$-axis, which is consistent with previous tight-binding calculations \cite{Mellinger2020}. 
On the other hand, those of Co$^{2+}$ at the $O_h$ site and Ni$^{2+}$ at the $T_d$ site are preferentially orientated in the $ab$-plane.
$\Delta E(\theta)$ of Ni$^{3+}$ at the $T_d$ site has a large $K_{u2}$ with a minimum of approximately 45 and 135 degree; therefore, it could be the origin of the ECMA.

Next, we attempted to deduce the total MA by considering the site distribution. 
Because the positive $K_{u2}$ possibly originates from Ni$^{3+}$ at the $T_d$ site, we need to introduce the anti-site disorders of Ni$^{3+}$ at the $T_d$ sites in our calculation.  
In fact, according to recent experimental research \cite{Shen2020}, approximately 10-20\% of the anti-site disorders are present at the $T_d$ site in the NCO films regardless of the growth conditions.  
To estimate the total $\Delta E(\theta)$, we assumed the cation distributions listed in Table \ref{distribution}. 
The distribution satisfies the assumption that the ratio of atoms is $\mathrm{Ni:Co:O=1:2:4}$, and the total valence becomes zero.
\begin{table}[htb]
\caption{Assumed cation distribution of ions for total MA calculation. }
\label{distribution}
\begin{ruledtabular}
\begin{tabular}{c|ccc|ccc}
Site              & \multicolumn{3}{c|}{$T_d$}    & \multicolumn{3}{c}{$O_h$}  \\ \hline
Ion               & Co$^{2+}$     & Co$^{3+}$     & Ni$^{3+}$ & Co$^{3+}$     & Ni$^{2+}$ & Ni$^{3+}$    \\ 
$S$              & 3/2           & 2        & 3/2    & 0     & 1     & 1/2    \\ 
Occupancy         & 0.85          &   0.0    & 0.15   & 1.15   & 0.15  & 0.70 \\
\end{tabular}
\end{ruledtabular}
\end{table}

Figure \ref{Cluster} (c) shows the results of the calculation of the total MA that takes into account the cation distribution in Table \ref{distribution}.
By fitting the total $\Delta E(\theta)$ using Eq. (1), we obtain $K_{u1} = -1.34$ MJ/m$^3$ ($-0.56$ meV/f.u.), and $K_{u2} = 1.42$ MJ/m$^3$(0.59 meV/f.u.). 
Because $K^{\mathrm{eff}}_{u1} < 0$ and $K_{u2} > 0$, the calculated result qualitatively reproduces the ECMA.
Note that both $K_{u1} = K^{\mathrm{eff}}_{u1} + \mu_0 M_0^2/2$ and $K_{u2}$ are in relatively good agreement with the experimentally obtained values, and especially closely approximate those for Sample $P_{10}$, as in Table  \ref{Summary}. 
According to the site/ion dependencies of $\Delta E$ shown in Fig. \ref{Cluster} (a) and (b), the small amount of Ni$^{3+}$ that occupies the $T_d$ site is a key component responsible for the ECMA.

\section{Conclusion}
We investigated the lattice distortions and magnetic properties of epitaxial NCO films grown on MAO(001) substrates by magnetron sputtering and PLD. 
Although the NCO films exhibited PMA at room temperature, the magneto-torque experiments revealed that the NCO films exhibited ECMA below $T_{\mathrm{SR}}$, regardless of the preparation technique.  
The observed spin reorientation seemed to occur continuously at $T_{\mathrm{SR}}$, and the cone angle increased with decreasing temperature. 
$T_{\mathrm{SR}}$ and $T_C$ are strongly dependent on the conditions under which the film was grown, suggesting that these conditions affect the distribution of the different cations.
The proposed electron theory based on the cluster model successfully explains the origin of the ECMA, which is attributed to the Ni$^{3+}$ that occupy the $T_d$ site.  
The results of both the experimental and cluster model calculations suggest that the ECMA is the result of the cation anti-site distribution of Ni$^{3+}$ associated with the growth conditions of the thin film. 
Although we succeeded in reproducing the ECMA within the framework of the cluster model calculation, the reason for the change in the sign of $K_{u1}$ at $T_{\rm SR}$ remains to be solved. 
The mechanism of the spin reorientation transition could be  related to the mixed-valence nature of NCO.

\begin{acknowledgments}
This project is partly supported by the Japan Science and Technology Agency (JST) under collaborative research based on industrial demand ``High Performance Magnets: Towards Innovative Development of Next Generation magnets'' (JP-MJSK1415),  KAKENHI(19KK0104, 19H05816, 21H01810) , the Ibaraki University-Tsukuba Joint Coordination Fund, and the Tanigawa Fund Promotion of Thermal Technology.
This work was performed with the approval of the ``Photon Factory Program Advisory Committee'' (proposals No.2017G602 and No.2016S2-005). 
This work was also supported by the Japan Society for the Promotion of Science (JSPS) ``Core-to-Core program (A)'' Advanced Research Networks
by a grant from the Integrated Research Consortium on Chemical Sciences and the International Collaborative Research Program of the Institute for Chemical Research at Kyoto University from the Ministry of Education, Culture, Sports, Science and Technology (MEXT)'' of Japan.
H. K. acknowledges support in the form of a Grant-in-Aid for JSPS Fellows (20J10749).
H. K. thanks T. Oki for his cooperation.
\end{acknowledgments}

\bibliography{bibliography}

\end{document}